# First step towards Devil's Staircase in Spin Crossover materials

Elzbieta Trzop,[a]+ Daopeng Zhang,[b,c]+ Lucia Piñeiro-Lopez,[b] Francisco J. Valverde-Muñoz,[b] M. Carmen Muñoz,[d] Lukas Palatinus,[e] Laurent Guerin,[a] Hervé Cailleau,[a] Jose Antonio Real*[b] and Eric Collet*[a]

## COMMUNICATION

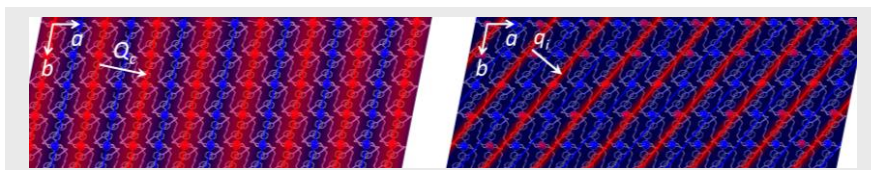

Periodic and aperiodic spin-state concentration waves form during Devil's staircase-type spin-crossover in the new bimetallic 2D coordination polymer $\{Fe[(Hg(SCN)_3)_2](4,4'bipy)_2\}_n$.

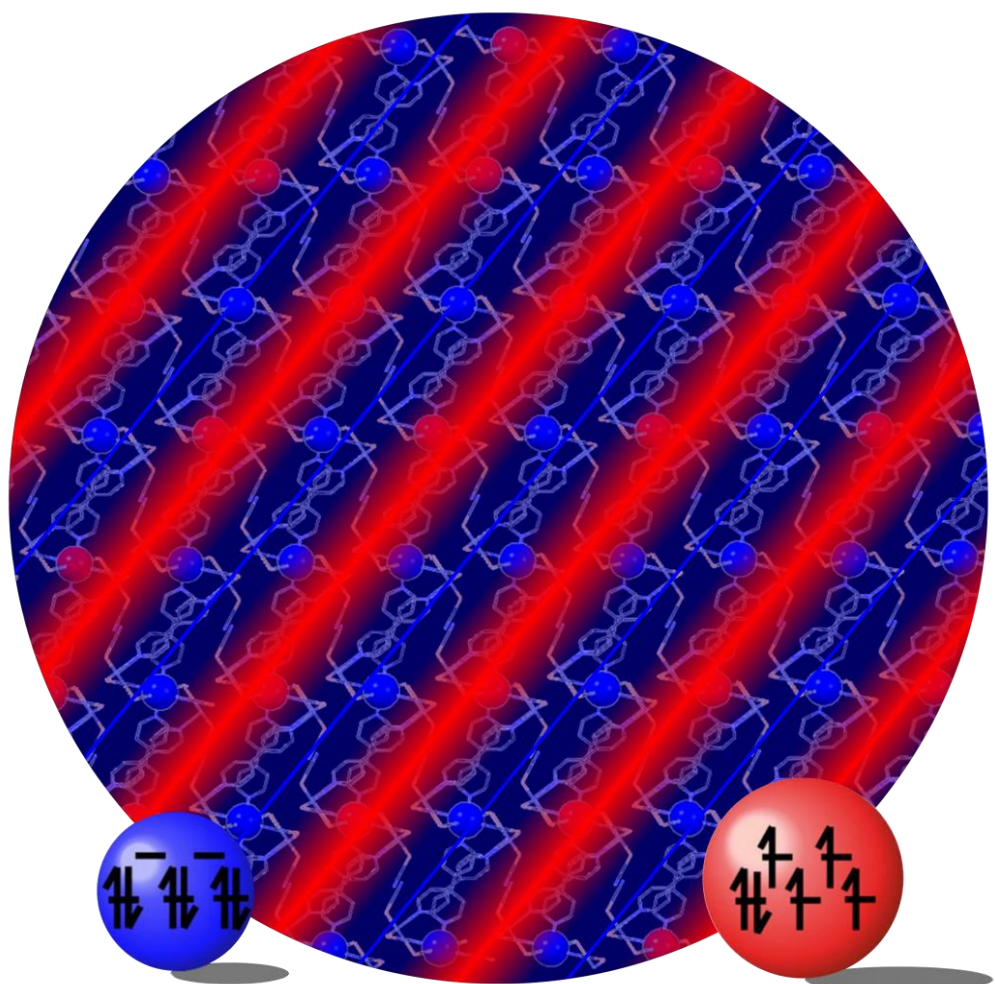

*Multi-step spin-crossover. In their Communication, E. Collet et al show that spin-state concentration waves form during multi-step spin-crossover. The formation of these different long-range spatially ordered structures of molecules in low or high spin state show some similarity with Devil's staircase-type phase transitions.*

# First step towards Devil's Staircase in Spin Crossover materials

Elzbieta Trzop,[a]+ Daopeng Zhang,[b,c]+ Lucia Piñeiro-Lopez,[b] Francisco J. Valverde-Muñoz,[b] M. Carmen Muñoz,[d] Lukas Palatinus,[e] Laurent Guerin,[a] Hervé Cailleau,[a] Jose Antonio Real*[b] and Eric Collet*[a]

**Abstract:** The unprecedented bimetallic 2D coordination polymer {Fe[(Hg(SCN)$_3$)$_2$](4,4'bipy)$_2$}$_n$ exhibits a thermal high-spin (HS)↔low-spin (LS) staircase-like conversion characterised by a multi-step dependence of the HS molar fraction $\gamma_{HS}$. Between the fully HS ($\gamma_{HS}$=1) and LS ($\gamma_{HS}$=0) phases, two steps associated with different ordering in terms of spin-state concentration waves (SSCW) appear. On the $\gamma_{HS}$≈0.5 step, a periodic SSCW forms with a ...HS-LS-HS-LS... sequence. On the $\gamma_{HS}$≈0.34 step, the 4D superspace crystallography structural refinement reveals an aperiodic SSCW, with a HS-LS sequence incommensurate with the molecular lattice. The formation of these different long-range spatially ordered structures of LS and HS states during the multi-step spin-crossover is discussed within framework of "Devil's staircase"-type transitions. Spatially modulated phases are known in various types of materials but are uniquely related to molecular HS/LS bistability here.

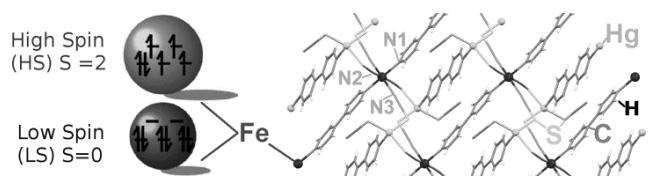

**Figure 1.** Electronic states of the Fe ion (left) and crystal structure of the 2D metallothiocyanate Fe(II)-Hg(II) SCO coordination polymer.

Spin-crossover (SCO) materials are made of bistable molecules able to switch from low (LS) to high spin (HS) states (Figure 1).[1] The LS to HS thermal conversion occurs when the entropy term ($T\Delta S$), favouring the HS state, compensates the enthalpy differences ($\Delta H$). The solely "ferro" elastic coupling, i.e. favouring energetically equal states between sites with higher (HS) or lower (LS) molecular volume, can drive cooperative LS to HS phase transformation. This is taken into account in Ising-like model through the temperature dependent on-site effective field $h=(-\Delta H+T\Delta S)/2$. The system follows in the universal Ising phase diagram an oblique line ($T, h(T)$), determining the thermal conversion of the HS molecular fraction $\gamma_{HS}$ from 0 to 1.[2] However, stepwise conversions have been reported in a vast variety of SCO materials made of Fe(II),[3] Fe(III),[4] Co(II),[5] or Mn(III)[6] ions, with the establishment of different HS-LS periodic order on the steps. This results from competing "ferro" and "antiferro" interactions.[7]

The ANNNI model, Anisotropic Next Nearest Neighbours Interactions, is an extension of the Ising model, which explains the formation of such modulated commensurate and incommensurate structures in crystals.[8] The concept of Devil's staircase was then theoretically introduced to explain the rich sequence of steps during the transformation of a system, as the on-site field $h$ (or $T$) changes (Figure 2).[9] Each step corresponds to a long-period commensurate structure, whose spatial modulation of states (represented by + or −) is associated with a wavevector $Q$ locked onto a rational number $n/m$. Between any two steps there is an infinity of steps, since between any two rational numbers there is an infinity of rational numbers. That's the reason why this sequence of phases is called the Devil's staircase. It appears in a vast variety of systems with states + or − of different nature, including charge or spin density waves, atomic concentration waves in alloys, ferroelectric materials or spin-valve systems. Such nanostructures open the possibility to create completely new electronic functionalities for technological applications.[10] Here we report on the incomplete Devil's staircase during the SCO of the new metallothiocyanate Fe(II)-Hg(II) 2D coordination polymer {Fe[(Hg(SCN)$_3$)$_2$](4,4'bipy)$_2$}$_n$ (**1**) as commensurate and incommensurate spin-state concentration waves, with different spatial sequences of HS (+) and LS (−) states, form on the steps.

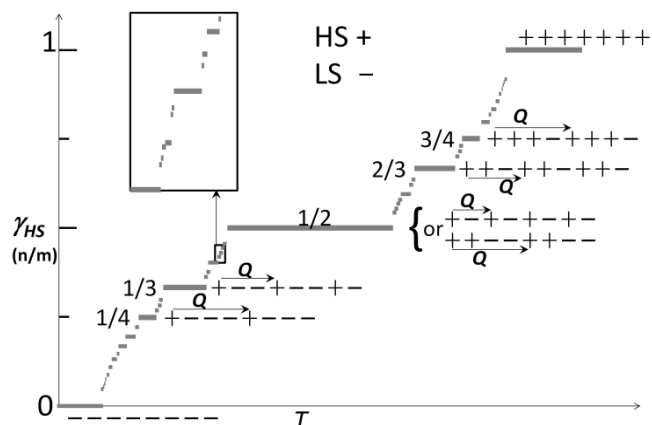

**Figure 2.** Theoretical Devil's staircase (modified from ref. 9d). For SCO materials the steps with $T$ (or $H$) correspond to fractions $\gamma_{HS}$=n/m of HS (+) molecules forming a periodic sequence with LS (−) molecules along the modulation vector $Q$. Structure of simplest (thick) steps are represented.

[a] Pr E. Collet, Dr E. Trzop, Dr L. Guérin and Pr H. Cailleau
Institut de Physique de Rennes, Université de Rennes 1,
UMR UR1-CNRS 6251, 35000 Rennes, France.
E-mail: eric.collet@univ-rennes1.fr

[b] Dr D. Zhang, Dr L. Piñeiro-López, Dr F.J. Valverde-Muñoz, Pr J.A. Real, Instituto de Ciencia Molecular (ICMol), Universidad de Valencia, C/ Catedrático José Beltrán Martínez 2, 46980 Paterna, Valencia, Spain
E-mail: Jose.A.Real@uv.es

[c] Dr D. Zhang
College of Chemical Engineering, Shandong University of Technology, Zibo 255049, China.

[d] Pr M. Carmen Muñoz
Departamento de Física Aplicada, Universitat Politècnica de València. Camino de Vera s/n, E-46022, Valencia, Spain

[e] Dr L. Palatinus
Department of Structure Analysis, Institute of Physics of Academy of Sciences of Czech Republic, Cz-182 21 Prague.

[+] These authors contributed equally to this work.

Supporting information for this article is given via a link at the end of the document.

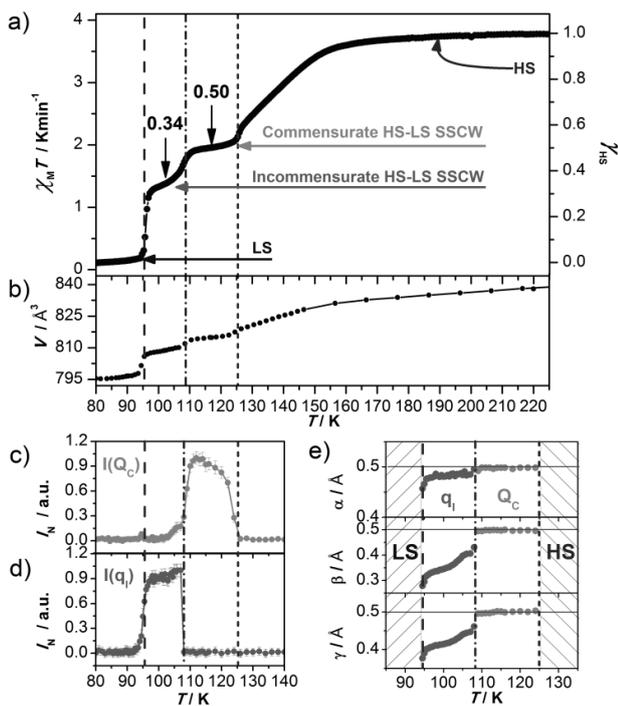

**Figure 3.** a) Temperature dependence of $\chi_M T$, scaled to $\gamma_{HS}$, measured at 1 K min$^{-1}$ cooling rate. b) Unit cell volume given in the **a, b, c** cell of the HS and LS phases. c) Intensity of superstructure peaks indexed $Q_c=1/2a^*+1/2b^*+1/2c^*$ in the HS **a\*, b\*, c\*** lattice. d) Intensity of the satellites reflections, indexed $q_i \approx \alpha a^* + \beta b^* + \gamma c^*$. e) Evolution of the coordinates $\alpha$, $\beta$ and $\gamma$ of $Q_c$ and $q_i$.

**1** was synthesized from self-assembly of Fe(II), 4,4'-bipyridine (4,4'-bipy) and [Hg(SCN)$_4$]$^{2-}$ in water/methanol solutions (Supporting Information). A fragment of the structure is shown in Figure 1. The Fe sites show bistability between LS (S=0) and HS (S=2) states. Its $\gamma_{HS}$-vs-T conversion was characterized by the evolution of the $\chi_M T$ product ($\chi_M$=magnetic susceptibility, T temperature) shown in Figure 3a. Above ≈180 K, $\chi_M T \approx 3.67$ cm$^3$ K mol$^{-1}$ is practically constant and consistent with a fully Fe$^{II}$ HS ($\gamma_{HS}$=1) phase. Below ≈95 K, $\chi_M T$ characterises a fully Fe$^{II}$ LS phase ($\gamma_{HS}$=0). $\gamma_{HS}$-vs-T conversion plot shows a multi-step conversion: in addition to the HS and LS phases, two other steps are observed. One in the 108-125 K interval, where $\gamma_{HS} \approx 0.5$ and one in the 96-108 K interval around $\gamma_{HS} \approx 0.34$. Single crystal X-ray diffraction data have been recorded for different temperatures[11] (Supporting Information) and the unit cell volume evolution confirms the steps observed in the magnetic measurement (Figure 3b). The HS structure (220 K, Table S1) is triclinic P$\bar{1}$ and constituted of slightly distorted centrosymmetric octahedral [Fe$^{II}$N$_6$] sites, axially connected through 4,4'bipy ligands (through N1) (Figure 1). The equatorial positions are occupied by the N2 and N3 atoms of two crystallographically unique NCS-groups. Their S atoms participate in strongly distorted tetrahedral [Hg$^{II}$S$_3$N] coordination sites. The dimeric building blocks {[Hg$^{II}$(SCN)$_3$]$_2$($\mu$-4,4'-bipy)} link consecutive "Fe-4,4'bipy" chains. At 220 K, all Fe$^{II}$ sites are crystallographically equivalent. The average Fe-N bond length <Fe-N> = 2.158(3) Å is typical of the HS state, in agreement with magnetic data. At 90 K, all Fe$^{II}$ sites are also crystallographically equivalent in a similar (**a b c**) cell and <Fe-N> = 1.962(3) Å is typical of the LS state (Table S1 & S4).

The <Fe-N> contraction contributes strongly to the lattice deformation in these polymeric SCO materials.[2f]

A symmetry breaking occurs on the 108-125 K plateau. It corresponds to a doubling along (**a+b+c**) and is characterized by the appearance of superstructure Bragg peaks at $Q_c=\frac{1}{2}a^*+\frac{1}{2}b^*+\frac{1}{2}c^*$ (Figure S1). The temperature dependence of their intensity (Figure 3c) indicates that this phase is surrounded by two first-order phase transitions around 125 K and 108 K, in agreement with magnetic data. The structural refinement at 117K (Table S2, S4) in the doubled cell reveals two different Fe sites: site 1 with <Fe1-N>≈1.962(3) Å is mainly LS and site 2 with <Fe2-N>≈2.146(3) Å is mainly HS. The HS fraction on each site can be estimated from its linear variation with <Fe-N>.[12] We estimate (Table S5) $\gamma_{HS} \approx 0.05$ on site 1 and $\gamma_{HS} \approx 0.95$ on site 2. The crystal structure consists then of mainly HS and LS stripes alternating in a ...HS-LS-HS-LS... sequence along $Q_c$. This can be described as a spin-state concentration wave (SSCW),[8b] which is commensurate with the initial (**a,b,c**) lattice (Figure 4a). It corresponds to the +−+−+− sequence on the 1/2 step of the Devil's staircase in figure 2. This spatial modulation of $\gamma_{HS}(\mathbf{r})$, translating through the modulation of <Fe-N>, is schematically represented by a wave in Figure 4a related to the modulation of the probability for a crystalline sites in **r** to be HS:

$$\gamma_{HS}^c(\mathbf{r}) = \gamma_{HS}^c + (\eta^c/2) \times \cos(\mathbf{Q}_c \cdot \mathbf{r}).$$

$\gamma_{HS}^c \approx 0.5$ can be estimated from <Fe-N>=2.058 Å averaged between sites 1 and 2 (Table S5), in agreement with magnetic data. $\eta^c \approx 0.9$ is the amplitude of the wave, related to the different probabilities for sites 1 or 2 to populate the HS state. In the ANNNI model,[9d] the 1/2 step is the largest one (Figure 2) and consequently the most likely to be observed and HS-LS-HS-LS or HS-HS-LS-LS sequences were reported in different SCO materials.[3,8] Other steps at $\gamma_{HS}$=1/3 (LS-LS-HS) or $\gamma_{HS}$=2/3 (HS-HS-LS) were also reported in the literature and characterized by a unit cell tripling.[13] The relative stability of the steps is governed by the subtle balance of the interactions and temperature and that's the reason why only a limited number of rational steps have been experimentally reported in SCO materials, including also 1/4, 3/4.[3-8,14] Since we observe here another step at $\gamma_{HS} \approx 0.34$, which is close to 1/3, one wonders about the nature of the order on this step. The symmetry breaking on the 96-108 K step does not correspond to a unit cell tripling. Different type of Bragg peaks appear, which cannot be indexed with the three vectors basis of the reciprocal lattice (Figure S1). A fourth vector $q_i$ is required for indexing the scattering vector **Q** of the peaks:

$\mathbf{Q}$=h$\mathbf{a}^*$+k$\mathbf{b}^*$+l$\mathbf{c}^*$+m$\mathbf{q}_i$. h, k, l & m $\in \mathbb{N}$ and $\mathbf{q}_i$=$\alpha\mathbf{a}^*$+$\beta\mathbf{b}^*$+$\gamma\mathbf{c}^*$.

At 102 K $\alpha \approx 0.48(1)$, $\beta \approx 0.35(1)$ and $\gamma \approx 0.42(1)$ cannot be expressed as simple fractions. The structure is therefore no more 3D periodic, but incommensurately modulated with respect to the initial (**a, b, c**) lattice.[15,16] It belongs to the 4D superspace group P$\bar{1}$($\alpha,\beta,\gamma$)0 (SSG number 2.1.1.1).[17] The discontinuous changes of the satellites' intensities (Figure 3d) around 96 and 108 K, concomitant with $\chi_M T$ and V jumps, underline the first order nature of the transitions delimiting the stability region of the incommensurate phase. This incommensurate structure was refined in a (3+1)-dimensional superspace[14] (Table S3), where atomic coordinates are described by their positions in the average 3D unit cell plus the modulation functions of these coordinates along the fourth dimension coordinate $x_4$ (Figure S2), consequently modulating the Fe-N distances (Figure 4c). The in-phase modulation of these bonds along $x_4$ is the direct proof of the formation of a SSCW along ***q_i***, incommensurate with the average 3D periodic crystal lattice, as observed in another SCO material.[16] The in-phase modulation of these bonds along $x_4$ is the direct proof of the formation of a SSCW along ***q_i***, incommensurate with the average 3D periodic crystal lattice, as observed in another SCO material.[16] The average modulation of the 6 Fe-N bonds indicates an almost sinusoidal modulation function of $\gamma_{HS}$: $\gamma_{HS}^i(r) = \gamma_{HS}^i + (\eta^i/2) \times \cos(q_i \cdot r)$ (Figure 4c) with $\gamma_{HS}^i$=0.34 and $\eta^i$=0.85 (Table S4 & S5). The HS fraction modulation along $x_4$, with <Fe-N> > 2.058 Å fits with $\gamma_{HS}^i$=0.34. The modulation vector ***q_i*** of $\gamma_{HS}^i(r)$, forming HS and LS stripes, is incommensurate with the (***a,b,c***) lattice. This 4D structure can then be projected in the physical 3D space to reveal the aperiodic spatial distribution of LS and HS states (Figure 4b).

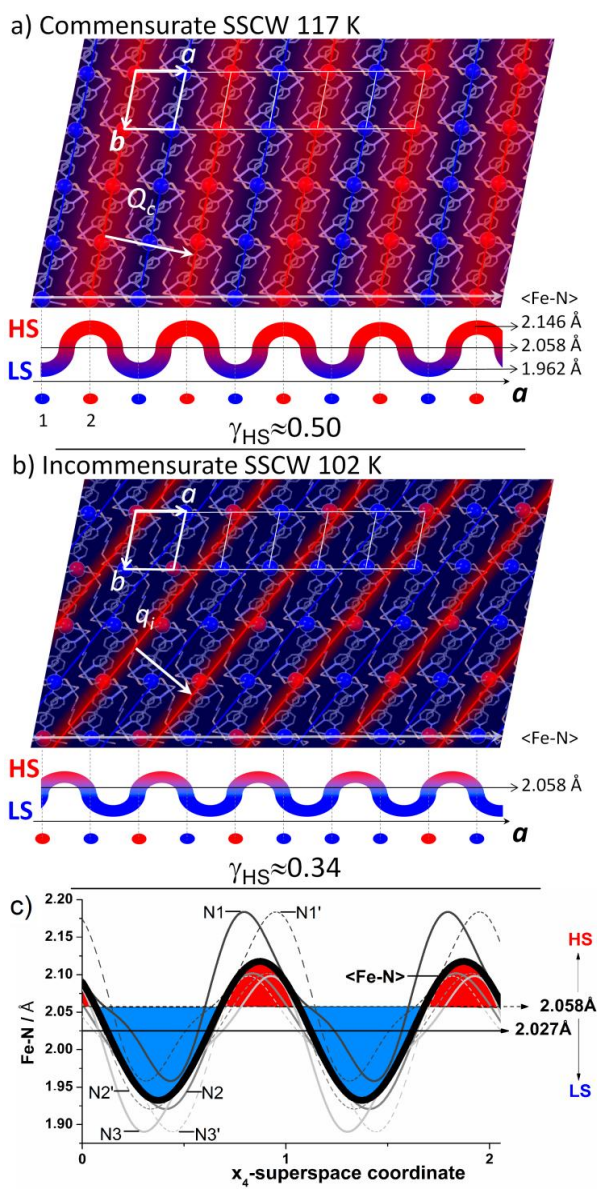

**Figure 4.** a) Striped structure sliced in the (***a,b***) plane of the SSCW forming along ***Q_c***. It is commensurate with the (***a,b,c***) lattice (white) and made of mainly LS (blue, <Fe-N>=1.962 Å) and HS (red, <Fe-N>=2.146 Å) sites alternating in the LS-HS-LS-HS sequence around <Fe-N>=2.058 Å ($\gamma_{HS}^c$≈0.5). b) Slice in the (***a,b***) plane of the aperiodic projected structure, forming an incommensurate SSCW with HS and LS stripes along ***q_i*** (projected). c) Modulation of the Fe-N bonds along the superspace coordinate $x_4$ with an almost sinusoidal modulation of <Fe-N> (thick line) around ≈2.027 Å corresponding to $\gamma_{HS}^i$≈0.34. The fraction of the Fe-N modulation larger than ≈2.058 Å corresponding to mainly HS state is also close to $\gamma_{HS}^i$ ≈ 0.34. The schematic waves at the bottom of a) and b) show a slice of the spatial HS/LS modulation along the ***a*** axis (white lines at the bottom of the structures).

In the case of a theoretical Devil's staircase, the conversion from $\gamma_{HS}$=1 to $\gamma_{HS}$=0 can occur through every rational value n/m (Figure 2). Is this $\gamma_{HS}^i$=0.34 step really corresponding to an incommensurate phase, or is it simply a high order n/m commensurate phase? On this step ***q_i*** and $\gamma_{HS}$ continuously change (Figure 3e) and the SSCW is therefore not locked at a commensurate position of the lattice. This behaviour is characteristic of incommensurate structures,[14] contrary to the commensurate step $\gamma_{HS}$=1/2 where ***Q_c*** is locked at ***Q_c***=½***a*** *+½***b*** *+½***c*** * from 108 to 125 K.

This study can be discussed as a first step towards Devil's staircase related to SCO from the fully HS ($\gamma_{HS}$=1) to LS ($\gamma_{HS}$=0) phases, with two long-range ordered structures on the steps:
- a commensurate SSCW with $\gamma_{HS}$≈0.5 and a locked modulation vector ***Q_c*** generating a HS-LS-HS-LS sequence,
- an incommensurate SSCW with $\gamma_{HS}$≈0.34 and with modulation vector slightly evolving around ***q_i***=0.48***a*** *+0.35***b*** *+0.42***c*** *.

A true Devil's staircase involves many transitions, in which the wave vector changes by small amount and is commensurate at each step. The present multi-step process corresponds to an incomplete Devil's staircase as an incommensurate phase forms.[9c] It proves that 3D periodic ordering is not always the most stable one. The SSCW concept in Figure 2 and 4 can be used to describe the different commensurate steps observed in different SCO crystals.[3-6,13,14] It shows basic features similar to charge density, spin density or atomic concentration waves,[10] but the present electronic order is unusual since related to entropy driven molecular magnetic bistability.

## Acknowledgements

This work was supported by the Spanish Ministerio de Economia y Competitividad (MINECO), FEDER (CTQ2013-46275-P), Unidad de Excelencia MDM-2015-0538, the Generalitat Valenciana through PROMETEO/2012/049. L.P.L. and F.J.V.M. thank to the Universidad de Valencia and a MINECO for a predoctoral (FPI) grant. D.Z. thanks the Natural Science Foundation of China and China Scholarship Council. This work was supported by the Institut Universitaire de France, the National Research Agency (ANR-13-BS04-0002), Rennes Metropole and CNRS (Post-Doc funding of E.T.).

**Keywords:** Phase transitions • Spin crossover • Coordination Polymers • Devil's staircase • Aperiodicity

[1] M. Halcrow, Ed., *Spin-crossover materials*, Wiley, West Sussex, **2013**, ISBN 9781119998679.

[2] a) P. Chakraborty, C. Enachescu, A. Humair, L. Egger, T. Delgado, A. Tissot, L. Guénée, C. Besnard, R. Bronisz, A. Hauser, *Dalton Trans.* **2014**, *43*, 17786-17796; b) M. Buron-Le Cointe, J. Hébert, C. Baldé, N. Moisan, L. Toupet, P. Guionneau, J.-F. Létard, E. Freysz, H. Cailleau, E. Collet, *Phys. Rev. B* **2012**, *85*, 064114; c) C. Lochenie, W. Bauer, A.P. Railliet, S. Schlamp, Y. Garcia, B. Weber, *Inorg. Chem.* **2014**, *53*, 11563–11572; d) A. Tissot, R. Bertoni, E. Collet, L. Toupet, M.L. Boillot, *J. Mater. Chem.* **2011**, *21*, 18347-18353; e) B. Weber, W. Bauer, J.

Obel, *Angew. Chem., Int. Ed.* **2008**, *47*, 10098−10101; f) E. Collet, L. Henry, L. Piñeiro-López, L. Toupet, J.A. Real, *Curr. Inorg. Chem.* **2016**, *6*, 61-66; g) R. Bertoni, M. Lorenc, H. Cailleau, A. Tissot, J. Laisney, M.-L. Boillot, L. Stoleriu, A. Stancu, C. Enachescu, E. Collet, *Nature Mater.* **2016**, doi:10.1038/nmat4606.

[3] a) D. Chernyshov, M. Hostettler, K. W. Törnroos, H.-B. Bürgi, *Angew. Chem., Int. Ed.* **2003**, *42*, 3825-3830; b) S. Bonnet, G. Molnár, C.J. Sánchez, M.A. Siegler, A.L. Spek, A. Bousseksou, W.-T. Fu, P. Gamez, J. Reedijk, *Chem. Mater.* **2009**, *21*, 1123-1136; c) N. Bréfuel, H. Watanabe, L. Toupet, J. Come, N. Matsumoto, E. Collet, K. Tanaka, J.-P. Tuchagues, *Angew. Chem., Int. Ed.* **2009**, *48*, 9304-9307; d) N. Brefuel, E. Collet, H. Watanabe, M. Kojima, N. Matsumoto, L. Toupet, K. Tanaka, J.-P. Tuchagues, *Chem. Eur. J.* **2010**, *16*, 14060-14068; e) S. Bonnet, M. A. Siegler, J. S. Costa, G. Molnár, A. Bousseksou, A.L. Spek, P. Gamez, J. Reedijk, *Chem. Commun.* **2008**, 5619-5621; f) M. Buron-Le Cointe, N. Ould Moussa, E. Trzop, A. Moréac, G. Molnár, L. Toupet, A. Bousseksou, J.-F. Létard, G.S. Matouzenko, *Phys. Rev. B* **2010**, *82*, 214106; g) K. Nakano, S. Kawata, K. Yoneda, A. Fuyuhiro, T. Yagi, S. Nasu, S. Morimoto, S. Kaizaki, *Chem. Commun.* **2004**, 2892-2893; h) D.L. Reger, C.A. Little, V.G. Young Jr., M. Pink, *Inorg. Chem.* **2001**, *40*, 2870-2874; i) T. Sato, K. Nishi, S. Iijima, M. Kojima, N. Matsumoto, *Inorg. Chem.* **2009**, *48*, 7211-7229; j) M. Yamada, H. Hagiwara, H. Torigoe, N. Matsumoto, M. Kojima, F. Dahan, J.-P. Tuchagues, N. Re, S. Iijima, *Chem. Eur. J.* **2006**, *12*, 4536-4549; k) K. W. Törnroos, M. Hostettler, D. Chernyshov, B. Vangdal, H.-B. Bürgi, *Chem. Eur. J.* **2006**, *12*, 6207-6215; l) V.A. Money, C. Carbonera, J. Elhaiek, M.A. Halcrow, J.A.K. Howard, J.-F. Létard, *Chem. Eur. J.* **2007**, *13*, 5503-5514; m) J. Kusz, M. Nowak, P. Gütlich, *Eur. J. Inorg. Chem.* **2013**, 832-842; n) W. Hibbs, P.J. van Koningsbruggen, A.M. Arif, W.W. Shum, J.S. Miller, *Inorg. Chem.* **2003**, *42*, 5645-5653; o) A. Lennartson, A.D. Bond, S. Piligkos, C.J. McKenzie, *Angew. Chem., Int. Ed.* **2012**, *51*, 11049-11052; p) M. Nihei, H. Tahira, N. Takahashi, Y. Otake, Y. Yamamura, K. Saito, H. Oshio, *J. Am. Chem. Soc.* **2010**, *132*, 3553-3560; q) G. Agustí, M.C. Muñoz, A.B. Gaspar, J.A. Real, *Inorg. Chem.* **2008**, *47*, 2552 -2561; r) T. Kosone, C. Kanadani, T. Saito, T. Kitazawa, *Polyhedron* **2009**, *28*, 1930-1934; s) J.B. Lin, W. Xue, B.Y. Wang, J. Tao, W.X. Zhang, J.P. Zhang, X.M. Chen, *Inorg. Chem.* **2012**, *51*, 9423-9430; t) C.J. Adams, M.C. Muñoz, R.E. Waddington, J.A. Real, *Inorg. Chem.* **2011**, *50*, 10633-10642; u) V. Martinez, A.B. Gaspar, M. Carmen Munoz, G.V. Bukin, G. Levchenko, A. Real, *Chem. Eur. J.* **2009**, *15*, 10960-10971; y) G. Agustí, A.B. Gaspar, M.C. Muñoz, J.A. Real, *Inorg. Chem.* **2007**, *46*, 9646-9654.

[4] a) B.J.C. Vieira, J.T. Coutinho, I.C. Santos, L.C.J. Pereira, J.C. Waerenborgh, V. da Gama, *Inorg. Chem.* **2013**, *52*, 3845-3850; b) M. Griffin, S. Shakespeare, H.J. Shepherd, C.J. Harding, J.-F. Létard, C. Desplanches, A.E. Goeta, J.A.K. Howard, A.K. Powell, V. Mereacre, Y. Garcia, A.D. Naik, H. Müller–Bunz, G.G. Morgan, *Angew. Chem., Int. Ed.* **2011**, *50*, 896-900; c) Z.-Y. Li, J.-W. Dai, Y. Shiota, K. Yoshizawa, S. Kanegawa, O. Sato, *Chem. Eur. J.* **2013**, *19*, 12948-12952; d) K.D. Murnaghan, C. Carbonera, L. Toupet, M. Griffin, M.M. Dîrtu, C. Desplanches, Y. Garcia, E. Collet, J.-F. Létard, G.G. Morgan, *Chem. Eur. J.* **2014**, 20, 5613-5618; e) D.J. Harding, W. Phonsri, P. Harding, K.S. Murray, B. Moubaraki, G.N.L. Jameson, *Dalton Trans.* **2015**, *44*, 15079-15082.

[5] a) M.G. Cowan, J. Olguín, S. Narayanaswamy, J.L. Tallon, S. Brooker, *J. Am. Chem. Soc.* **2012**, *134*, 2892-2894; b) S. Hayami, T. Komatsu, T. Shimizu, H. Kamihata, Y.H. Lee, *Coord. Chem. Rev.* **2011**, *255*, 1981-1990; c) S. Hayami, K. Murata, D. Urakami, Y. Kojima, M. Akita, K. Inoue, *Chem. Commun.* **2008**, 6510-6512; d) S. Hayami, Y. Shigeyoshi, M. Akita, K. Inoue, K. Kato, K. Osaka, M. Takata, R. Kawajiri, T. Mitani, Y. Maeda, *Angew. Chem., Int. Ed.* **2005**, *44*, 4899-4903.

[6] A.J. Fitzpatrick, E. Trzop, E. Müller-Bunz, M. Dîrtu, Y. Garcia, E. Collet, G.G.Morgan, *Chem. Commun.* **2015**, *51*, 17540- 17543.

[7] a) A. Bousseksou, F. Varret, J. Nasser, *J. Phys. I* **1993**, *3*, 1463-1473; b) D. Chernyshov, H.-B. Burgi, M. Hostettler, K.W. Törnroos, *Phys. Rev. B* **2004**, *70*, 094116; c) M. Nishino, K. Boukheddaden, S. Miyashita, F. Varret, *Phys. Rev. B* **2003**, *68*, 224402.

[8] a) K. Boukheddaden, J. Linares, R. Tanasa, C. Chong, *J. Phys.: Condens. Matter* **2007**, *19*, 106201; b) H. Watanabe, K. Tanaka, N. Bréfuel, H. Cailleau, J.-F Létard, S. Ravy, P. Fertey, M. Nishino, S. Miyashita, E. Collet, *Phys. Rev. B* **2016**, *93*, 014419.

[9] a) P. Bak, J. von Boehm, *Phys. Rev.* **1980**, *21*, 5297; b) M.E. Fisher, W. Selke, *Phys. Rev. Lett.* **1980**, *44*, 1502; c) S. Aubry, *J. Physique* **1983**, *44*, 147-162; d) P. Bak, R. Bruinsma, *Phys. Rev. Lett.* **1982**, 49, 249-251.

[10] a) M. Gao, C. Lu, H. Jean–Ruel, L.C. Liu, A. Marx, K. Onda, S. Koshihara, Y. Nakano, X. Shao, T. Hiramatsu, G. Saito, H. Yamochi, R.R. Cooney, G. Moriena, G. Sciaini, R.J.D. Miller, *Nature* **2013**, *496*, 343-346; b) P. Monceau, *Advances in Physics* **2012**, *61*, 325-581; c) K. Ohwada, Y. Fujii, N. Takesue, M. Isobe, Y. Ueda, H. Nakao, Y. Wakabayashi, Y. Murakami, K. Ito, Y. Amemiya, H. Fujihisa, K. Aoki, T. Shobu, Y. Noda, N. Ikeda, *Phys. Rev. Lett.* **2001**, *87*, 086402; d) K. Machida, M. Nakano, *J. Phys. Soc. Jpn.* **1990**, *59*, 4223-4226; e) A. Murakami, Y. Tsunoda, *Phys. Rev. B* **2000**, *61*, 5998; f) M. Hupalo, J. Schmalian, M.C. Tringides, *Phys. Rev. Lett.* **2003**, *90*, 216106; g) T. Isozaki, T. Fujikawa, H. Takezoe, A. Fukuda, T. Hagiwara, Y. Suzuki, I. Kawamura, *Phys. Rev. B* **1993**, *48*, 13439-13450; h) T. Matsuda, S. Partzsch, T. Tsuyama, E. Schierle, E. Weschke, J. Geck, T. Saito, S. Ishiwata, Y. Tokura, H. Wadati, *Phys. Rev. Lett.* **2015**, *114*, 236403.

[11] Details on crystallographic data for 220 K (CCDC-1457783), 90 K (CCDC-1457781), 117 K (CCDC-1457782) and 102 K (CCDC-1457785) can be found in supporting information.

[12] E. Collet, M.L. Boillot, J. Hébert, N. Moisan, M. Servol, M. Lorenc, L. Toupet, M. Buron-Le Cointe, A. Tissot, J. Sainton, *Acta Cryst. B* **2009**, *65*, 474-480.

[13] a) S. Pillet, E.E. Bendeif, S. Bonnet, H.J. Shepherd, P. Guionneau, *Phys. Rev. B* **2012**, *86*, 064106; b) J. Luan, J. Zhou, Z. Liu, B. Zhu, H. Wang, X. Bao, W. Liu, M.-L. Tong, G. Peng, H. Peng, L. Salmon, A. Bousseksou, *Inorg. Chem.* **2015**, *54*, 5145–5147; c) G. Agustí, A.B. Gaspar, M.C. Muñoz, P. Lacroix, J.A. Real, *Aust. J. Chem.* **2009**, *62*, 1155-1165; d) Z.-Y. Li, H. Ohtsu, T. Kojima, J.-W. Dai, T. Yoshida, B.K. Breedlove, W.-X. Zhang, H. Iguchi, O. Sato, M. Kawano, M. Yamashita *Angew. Chem.* **2016**, *128*, 1–7.

[14] a) N.F. Sciortino, K.R. Scherl-Gruenwald, G. Chastanet, G.J. Halder, K.W. Chapman, J.-F. Létard, C.J. Kepert, *Angew. Chem., Int. Ed.* **2012**, *51*, 10154–10158; b) M. Nihei, H. Tahira, N. Takahashi, Y. Otake, Y. Yamamura, K. Saito, H. Oshio, *J. Am. Chem. Soc.* **2010**, *132*, 3553-3560K; c) D. Murnaghan, C. Carbonera, L. Toupet, M. Griffin, M.M. Ditru, C. Desplanches, Y. Garcia, E. Collet, J.-F. Létard, G. Morgan, *Chem. Eur. J.* **2014**, *20*, 5613-5618; d) N. Brefuel, E. Collet, H. Watanabe, M. Kojima, N. Matsumoto, L. Toupet, K. Tanaka, J.-P. Tuchagues, *Chem. Eur. J.* **2010**, *16*, 14060-14068.

[15] a) T. Janssen, G. Chapuis, M. de Boissieu, *Aperiodic Crystals: From Modulated Phases to Quasicrystals*, Oxford Univ. Press, Oxford, **2007**; b) S. van Smaalen. *Incommensurate Crystallography*, Oxford Univ. Press, Oxford, **2007**; c) A. Janner, T. Janssen, J.C. Toledano, *Lecture Notes in Physics* **1984**, *201*, 364-396.

[16] E. Collet, H. Watanabe, N. Bréfuel, L. Palatinus, L. Roudaut, L. Toupet, K. Tanaka, J.P. Tuchagues, P. Fertey, S. Ravy, B. Toudic, H. Cailleau, *Phys. Rev. Lett.* **2012**, *109*, 257206.

[17] H.T. Stokes, B.J. Campbell, S. van Smaalen, *Acta Cryst. A* **2011**, *67*, 45-55.